\documentclass[pdftex,twocolumn,epjc3]{svjour3}
\RequirePackage[T1]{fontenc}
\smartqed  

\RequirePackage{graphicx,amssymb}
\RequirePackage{mathptmx}      
\RequirePackage{flushend}
\RequirePackage[numbers,sort&compress]{natbib}
\RequirePackage[colorlinks,citecolor=blue,urlcolor=blue,linkcolor=blue]{hyperref}
\journalname{Eur. Phys. J. C}

\def\be{\begin{equation}}
\def\ee{\end{equation}}
\newcommand{\Herwig}{\small \textsc{Herwig++} \normalsize}
\newcommand{\madgraph}{\small \textsc{MadGraph} \normalsize}
\newcommand{\powhegbox}{\small \textsc{PowhegBox} \normalsize}
\newcommand{\powheg}{\small \textsc{Powheg} \normalsize}

\newcommand{\fastjet}{\small \textsc{FastJet} \normalsize}
\newcommand{\GeV}{\,\mathrm{GeV}}
\newcommand{\ie}{\textit{i.e.}}

\begin{document}
\title{Studying the sensitivity of monotop probes to compressed supersymmetric scenarios at the LHC}

\author{Benjamin Fuks\thanksref{e1,addr1,addr2}
  \and
  Peter Richardson\thanksref{e2,addr3}
  \and
  Alexandra Wilcock\thanksref{e3,addr3}}

\institute{CERN, PH-TH, CH-1211 Geneva 23, Switzerland\label{addr1}
  \and
  Institut Pluridisciplinaire Hubert Curien/D\'epartement
  Recherches Subatomiques, Universit\'e de Strasbourg/CNRS-IN2P3,
  23 rue du Loess, F-67037 Strasbourg, France \label{addr2}
  \and
  Institute for Particle Physics Phenomenology, University of
  Durham, Durham, DH1 3LE, United Kingdom\label{addr3}}

\thankstext{e1}{e-mail: benjamin.fuks@cern.ch}
\thankstext{e2}{e-mail: peter.richardson@durham.ac.uk}
\thankstext{e3}{e-mail: a.h.wilcock@durham.ac.uk}

\date{Received: date / Accepted: date}

\maketitle

\begin{abstract}
We investigate the sensitivity of the Large Hadron Collider to
supersymmetric setups using monotop probes in which the signal is a single
top quark produced in association with missing transverse energy. Our
prospective study relies on Monte Carlo simulations of $300~\mathrm{fb}^{-1}$ of
proton-proton collisions at a centre-of-mass energy of 14~TeV
and considers both leptonic and hadronic monotop decays. 
 We present analysis strategies sensitive to regions of the supersymmetric parameter
space which feature small superparticle mass splittings and illustrate their strengths
in the context of a particular set of benchmark scenarios.  Finally, we compare the regions of parameter space expected to be accessible with monotops probes during the next run of the LHC to the reach of more traditional search strategies employed by the ATLAS and CMS collaborations, where available.

\end{abstract}



\flushbottom

\section{Introduction}
After more than fifty years of experimental tests, the Standard Model
has proven to be a successful theory of elementary particles and
their interactions. While it consistently predicts
most existing high-energy physics data, it additionally includes a set of
conceptual problems for which it does not provide a satisfactory answer.
It is therefore widely believed to be the low-energy limit of a more fundamental
theory, weak scale supersymmetry~\cite{Nilles:1983ge,Haber:1984rc} being one of the
most popular and studied candidates.
By associating a partner of opposite statistics with each of the
Standard Model degrees of freedom, supersymmetric theories feature a way
to unify the Poincar\'e symmetry with the internal gauge symmetries and
provide an elegant solution to the hierarchy problem, amongst other appealing
theoretical features.

Since, so far, no hint for new physics has been clearly identified,
the superpartners of the Standard Model particles are constrained
to lie at higher and higher scales~\cite{atlassusy,cmssusy}. Most of these
bounds can however be evaded for compressed supersymmetric models
where the ensemble of states accessible at the Large Hadron Collider (LHC) exhibit
small mass differences. In particular, both the ATLAS and CMS experiments have
been found to be insensitive to scenarios with mass gaps of about 10~GeV or less
between the strongly interacting superparticles and the lightest superpartner.
In these cases, pair-produced squarks and gluinos decay into missing
energy carried by the lightest supersymmetric particle and
leptons and/or jets too soft to reach the typical
trigger thresholds of the LHC experiments. Moreover, the expected amount of
missing transverse energy is smaller, which implies first that the kinematic
quantities traditionally employed to reduce the Standard Model background are
less efficient and second that one cannot even rely solely on missing energy
triggers~\cite{atlasmet,cmsmet}. Classical search strategies
based on the presence of numerous jets and leptons and a large amount
of missing energy thus have poor sensitivity 
to compressed supersymmetric scenarios. Consequently, non-standard analyses
have been developed, making use for instance of monojet or monophoton
signatures~\cite{Alves:2010za,LeCompte:2011cn,LeCompte:2011fh,%
Alvarez:2012wf,Dreiner:2012gx,Bhattacherjee:2012mz,Dreiner:2012sh,%
Ghosh:2013qga,Belanger:2013oka,Dutta:2013gga,%
Bhattacherjee:2013wna,Schwaller:2013baa,Deppisch:2014aga}. They focus
on topologies where a superparticle pair is produced together with an
extra jet or photon that originates from initial-state radiation and can
further be used both for triggering and reducing the Standard Model background.

In this work, we explore a novel way of accessing the compressed regions of the
parameter space that relies on monotop probes, \ie, systems
comprised of missing transverse energy and a singly-produced top quark.
Mo\-no\-top sta\-tes are expected to be easily observable at the LHC for a
large range of new physics masses and couplings~\cite{Andrea:2011ws,%
Kamenik:2011nb,Wang:2011uxa,Fuks:2012im,Alvarez:2013jqa,Agram:2013wda,%
Boucheneb:2014wza}, although they have not been
experimentally found yet~\cite{Aaltonen:2012ek,CMS:2014hba,Khachatryan:2014uma,Aad:2014wza}.
We consider a simplified compressed supersymmetric scenario in which the
electroweak superpartners are ne\-glec\-ted with the exception of the lightest
neutralino.  Events describing the production and decay of a strong superpartner
pair in association with a top quark can manifest themselves via a monotop
signature when the decay of each superpartner gives rise to a small amount of
missing energy and soft objects.

The rest of this paper is organised as follows. In Section~\ref{sec:simulation},
we describe our technical setup for the Monte Carlo simulations of LHC
collisions at $\sqrt{s}=14$~TeV, both for the new physics signals and the
relevant sources of background. Our analysis strategy to extract a monotop
signal is detailed in Section~\ref{sec:selection} and the
results for specific benchmark scenarios are presented in
Section~\ref{sec:results}. Our conclusions are given in Section~\ref{sec:conclusions}.

\section{Technical setup for the Monte Carlo simulations} \label{sec:simulation}
\subsection{Signal simulation}\label{sec:simusignal}

In our simplified model framework, we first consider the production of a top quark in association with the lightest top squark $\tilde t_1$ and the gluino $\tilde g$, with the latter two decaying into the lightest neutralino $\tilde\chi_1^0$.  We study in detail the scenario in which the masses of the stop and gluino are similar and not significantly larger than that of the neutralino, such that additional Standard Model objects produced during the superparticle decays are soft and invisible. Monotop systems may also be produced in scenarios where the gluino is significantly heavier than the stop, which is again not much heavier than the neutralino.  In this case, monotop production relies on the associated production of a heavy gluino and a light top squark, the gluino decays into a top quark and a stop, while both stops are invisible.  Such a process occurs at tree-level when considering flavour-mixings of the up-type squarks that, for instance, occurs in minimally-flavor-violating supersymmetric models.  Production cross sections in such models are highly suppressed given that all flavour-violating effects are driven by the CKM matrix, however could be enhanced with non-minimal flavour violation~\cite{Bozzi:2007me,Fuks:2008ab,Fuks:2011dg}.  Nevertheless, we study only the first class of models.

In addition to the case in which a top squark and top quark are produced along with a gluino, we consider also the signal in which they are produced in association with the lightest neutralino.  Again, here we focus on the scenario in which the top squark is not significantly heavier than the neutralino.  For both signal scenarios, the stop is chosen to be a maximal admixture of the left-handed and right-handed stop gauge-eigenstates.  For the signal scenario in which a gluino is produced with a top squark and top quark, the lightest neutralino is assumed to be purely bino.  When instead a neutralino is produced directly, it is assumed to be predominantly higgsino\footnote{The neutralino mixing matrix is chosen such that the lighest neutralino is 99\% higgsino and 1\% bino.  In doing so, the production cross section is enhanced while still leaving all top squark decay modes accessible. In scenarios with a predominantly higgsino $\tilde{\chi}_1^0$, the lightest chargino and neutralino will be near degenerate in mass.  As such, the production of a top squark and bottom quark in association with the lightest chargino will lead to a ``mono $b$-jet'' signature that could be used to constrain the same region of parameter space as our monotop probe.  This signature provides an interesting alternative to monotop signals, however, its investigation goes beyond the scope of this work.}.  This enhances the production cross section as compared with the purely bino neutralino scenario.  Event generation for the hard scattering signal process relies on the implementation of the Minimal Supersymmetric Standard Model~\cite{Christensen:2009jx,Duhr:2011se,Degrande:2011ua,Alloul:2013bka} in the
\madgraph 5 program~\cite{Alwall:2014hca} that has been used to convolute hard matrix elements with the CTEQ6L1 set of parton density functions (PDF)~\cite{Pumplin:2002vw}. We describe in this way the production of a pair of strong superpartners in association with either a leptonically or a hadronically decaying top quark. 

The decays of the superparticles and matching of the parton-level hard events with a parton shower and hadronization infrastructure has been performed with the \Herwig~2.7 program~\cite{Bahr:2008pv,Bellm:2013lba}. The accessible decays modes of the final-state superparticles and associated branching ratios we\-re calculated internally by the \Herwig program.  No 4-body modes were considered and therefore in the highly compressed region with $m_{\tilde{t}_1} < m_{b}+m_{W}+m_{\tilde{\chi}_1^0}$ the dominant decay channel of the stop was\footnote{In the MSSM, the decay $\tilde{t}_1 \rightarrow c\, \tilde{\chi}_1^0 $ proceeds via a loop and CKM suppressed channel only.  Therefore, the width of the top squark is sufficiently small that hadronization occurs before the top squark decays.  However, the lifetime of the top squarks is still short when compared with detector time scales and so we assume the final observables do not differ significantly from the situation in which the top squarks decay promptly, before hadronization. } $\tilde{t}_1 \rightarrow c\, \tilde{\chi}_1^0 $.  In regions of phase space where $m_{\tilde{t}_1} > m_{b}+m_{W}+m_{\tilde{\chi}_1^0}$, the decay mode $\tilde{t}_1 \rightarrow W^+ b \tilde{\chi}_1^0$ was found to dominate.  For all mass scenarios considered in this study, the dominant gluino decay channel was $\tilde{g} \rightarrow q \, \bar{q} \, \tilde{\chi}_1^0$.  Finally, when simulating the production of a top squark, top quark and neutralino, a global normalization factor of $K=1.4$ has been applied to the total cross section.  This factor aims to account for the large next-to-leading order contributions to this process~\cite{Degrande:2014sta}.

Signal cross sections for the example scenarios of top squark, top quark and gluino production
with $(m_{\tilde{t}_1}\!=\!m_{\tilde{g}},m_{\tilde{\chi}_1^0})$
$= \left(200,190 \right)$~GeV and top squark, top quark and neutralino production with $(m_{\tilde{t}_1},m_{\tilde{\chi}_1^0})=\left(145,75 \right) $~GeV are given in Table~\ref{tab:sigma}. 
Comparing with the expected monojet cross sections of 12.5~pb (for $m_{\tilde{t}_1}=m_{\tilde{g}} = 200$~GeV after imposing a typical monojet requirement on the jet transverse momentum of $p_T>450$~GeV) and 0.7~pb (for $m_{\tilde{t}_1} = 145$~GeV again after imposing a requirement on the jet transverse momentum of $p_T>450$~GeV), one can expect these scenarios to be sensitive to both monotop and monojet probes.  However, the competitive cross section in the case of top quark, top squark and neutralino production is strongly affected by the composition of the neutralino.  In scenarios with a lightest neutralino that is predominately bino or wino, the cross section drops to 0.09~pb and 0.2~pb respectively when $(m_{\tilde{t}_1},m_{\tilde{\chi}_1^0})=\left(145,75 \right) $~GeV, with no selection criteria imposed.  In these cases, it is unlikely that monotop probes could provide comparable limits to those derived using monojet searches.

\subsection{Background processes}
Leptonically decaying monotop states yield event topologies comprised of one
hard lepton, one jet originating from the fragmentation of a $b$-quark and
missing transverse energy. As such, the main sources of background events consist of
the production of a $t \bar{t}$ pair where one of the top quarks decays
leptonically and the other one hadronically, as well as from the production of
a single-top quark in association with a $W$-boson where either the top quark or
the $W$-boson decays leptonically. We also consider extra background processes
expected to subdominantly contribute, namely the two other single-top production modes
and $W$-boson plus jets, $\gamma^*/Z$-boson plus jets and diboson production.

Turning to hadronically decaying monotops, the above signal final state is
altered with the hard lepton being replaced by a pair of hard jets. In this case,
background events are dominantly comprised of fully hadronic $t\bar{t}$ events,
$Z$-boson plus jets events in which the $Z$-boson decays invisibly as well as
$W$-boson plus light-jets events in which the $W$-boson decays leptonically but
where its decay products escape identification\footnote{The production of a
hadronically decaying $W$-boson plus-jets does not induce a significant background
contribution, owing to the small amount of expected missing energy.}.
Additionally, single-top, $W$-boson plus $b$-jets and diboson
production processes are also expected to contribute, in a subdominant way, to the
total number of monotop background events.

In the simulation of the Standard Model
backgrounds for both the leptonic and hadronic monotop analyses, QCD multijet production
processes have been neglected.  We instead assume the related background contributions will
be under good control after selection requirements such as
those detailed in Section~\ref{sec:selection} have been applied. Finally, all possible sources of
instrumental background are also ignored.  Consideration of these effects goes beyond the scope of
this work which does not aim to simulate any detector effect other than a $b$-tagging
efficiency (see Section~\ref{sec:selection}).

\renewcommand{\arraystretch}{1.2}
\begin{table*}[t]
\centering
\begin{tabular}{c||c||c|c||c|c}
  Process &  $\sigma [\mathrm{pb}]$ & $N_{\mathrm{event}}^{\mathrm{SRL1}}$ &  $N_{\mathrm{event}}^{\mathrm{SRL2}}$& $N_{\mathrm{event}}^{\mathrm{SRH1}}$ & $N_{\mathrm{event}}^{\mathrm{SRH2}}$ \\ \hline \hline
  $W ( \rightarrow l \nu ) $ + light-jets & $67453 $ &  $ \approx 0 $ & $ 3150 $ & $ 4500 $ & $ 9030 $ \\ 
  $\gamma^* / Z ( \rightarrow l \bar{l} ) $ + jets   & $26603  $ & $ \approx 0  $ & $ \approx 0$ & - & -\\ 
  $\gamma^* / Z ( \rightarrow \nu \bar{\nu} ) $ + jets & $12387  $ & $ \approx 0  $ & $\approx 0 $ & $ 23160  $ & $36390$ \\ 
  $t\bar{t}$  & $781 $ &  $  43230$ & $ 292500 $ & $ 35190   $ & $80040$\\ 
  Single top [$t$-channel] & $7320  $  &  $ 36.6   $& $4650  $ & $250.8 $ & $ 762 $\\ 
  Single top [$s$-channel] & $312  $  &  $ 6.3  $& $244.8 $ & $35.7$ & $ 75.3 $\\ 
  $tW$ production & $2313  $  &  $ 4890  $ & $42570  $ & $3480   $ & $7560$\\ 
  $Wb\bar{b}$ with $W \rightarrow l \nu$  & $3660$  &  $ \approx 0$& $549 $ & $134.1   $ & $158.7$\\ 
  Diboson & $158$ &  $ 31.5 $ & $ 268.8 $ & $205.5 $ & $315 $ \\  \hline
  Total background & $107834$ &  $ 48190 $ & $ 343900 $ & $ 66960 $ & $134330 $\\  \hline \hline
  $(m_{\tilde{g}}=m_{\tilde{t}_1},m_{\tilde{\chi}_1^0}) = (200, 190) \GeV $ & $ 2.54 $ & $8430  $ & $17280  $ & $12690 $ & $14700 $\\ 
  $\hspace{0.7cm}(m_{\tilde{t}_1},m_{\tilde{\chi}_1^0}) = (145, \,\,\,75)\, \GeV $ & $ 2.37 $ & $3180$ & $ 7773 $ & $6796   $ & $9840 $\\ 
\end{tabular}
\caption{Cross sections for the simulated background processes and two
  representative signal scenarios, including an NLO K-factor of 1.4 for the signal scenario in which the stop and top are produced in association with a neutralino. Also shown are the number of events,
  $N_{\mathrm{events}}$, surviving all selection criteria in the leptonic (SRL1,
  SRL2) and hadronic (SRH1, SRH2) signal regions. Results correspond to
  300~fb$^{-1}$ of LHC collisions at a centre-of-mass energy of 14~TeV.}
\label{tab:sigma}
\end{table*}
\renewcommand{\arraystretch}{1.0}

Parton-level hard events arising from the production of a top-antitop pair,
including top decays, have been simulated by convoluting next-to-leading order (NLO) matrix elements with the CTEQ6M parton density set~\cite{Pumplin:2002vw}
in the \powhegbox framework~\cite{Nason:2004rx,Frixione:2007vw,Alioli:2010xd,%
Frixione:2007nw}, and events have then been matched to \Herwig for parton
showering and hadronization\footnote{Owing to the angular-ordered nature of the  \textsc{Herwig++} parton shower, it is in principle necessary to apply a truncated parton showering algorithm to simulate emissions that have a smaller transverse momentum than those described by the NLO matrix elements but a larger value of the angular evolution parameter. However, the corresponding effects are typically small~\cite{Hamilton:2008pd} and so have been omitted.}.
The same machinery has been used to generate single-top
events~\cite{Alioli:2009je,Re:2010bp}, suppressing the doubly-resonant diagrams
related to the $tW$ mode following the prescription of Ref.~\cite{Frixione:2008yi},
as implemented in the \powhegbox.

In order to generate events describing the production of a $W$-boson with light-flavour jets $(u,d,s,c)$ and a $\gamma ^* / Z$-boson with both light- and heavy-flavour jets $(u,d,s,c,b)$, the {\sc Sherpa}~2.0~\cite{Gleisberg:2003xi,Gleisberg:2008ta} package has been used.
We have followed the MENLOPS prescription to match an event sample based on NLO matrix elements related to the production of a single gauge boson to leading-order (LO) samples describing the production of the same gauge boson with one and two extra jets~\cite{Hamilton:2010wh,Hoche:2010kg}. In all cases, the vector bosons have been forced to decay either leptonically or invisibly, including all three flavours of leptons, and matrix elements have been convoluted with the CTEQ6M PDF set. Moreover, the invariant masses of lepton pairs produced via a $\gamma^*/Z$-boson $s$-channel diagram have been required to exceed $10 \GeV$.

The production of a $W$-boson with heavy flavour jets has been simulated
separately using \madgraph 5 and its built-in Standard Model implementation.
We have generated LO matrix elements that have been convoluted with the LO set of parton densities CTEQ6L1~\cite{Pumplin:2002vw}. Parton-level events have been simulated including the leptonic decay of the $W$-boson and then showered and hadronized with \Herwig.

Finally, diboson production has been simulated at the NLO accuracy and matched
to the \Herwig parton shower using its built-in \powheg
implementation~\cite{Hamilton:2010mb}, the matrix elements having been convoluted with the CTEQ6M PDF set.

The total cross sections for all considered background processes are
shown in Table~\ref{tab:sigma}.

\section{Selection strategies} \label{sec:selection}
\subsection{Object reconstruction }
Objects used as inputs for the leptonic and hadronic monotop search strategies of
the next subsections are reconstructed as in typical single-top studies
performed by the ATLAS experiment (see, \textit{e.g.}, Ref.~\cite{Aad:2012ux}).
Electron (muon) candidates are required to have a
transverse momentum $p_T^\ell>10$~GeV, a pseudorapidity satisfying
$|\eta^\ell|<2.47$ (2.5) and they
must be isolated such that the
sum of the transverse momenta of all charged particles in a cone of
radius $\Delta R < 0.2$\footnote{The angular distance between two particles is defined
as $\Delta R = \sqrt{\Delta \phi^2 + \Delta \eta^2}$, $\Delta \phi$ and $\Delta \eta$
denoting their differences in the azimuthal angle with respect to the beam direction
and in pseudorapidity, respectively.} centered on the lepton is less than 10\% of its
transverse momentum.

Jets are reconstructed from all visible final-state particles with a
pseudorapidity satisfying \mbox{$|\eta^j| < 4.9$} by applying an anti-$k_T$ jet
algorithm~\cite{Cacciari:2008gp} with a radius parameter $R = 0.4$, as
implemented in the \fastjet program~\cite{Cacciari:2011ma}. We select
reconstructed jet candidates that do not overlap with candidate
electrons within a distance of $\Delta R < 0.2$, and with a transverse momentum $p^j_{T} >20$~GeV and
a pseudorapidity  \mbox{$|\eta^j| < 2.5$}. Any lepton candidate within
a distance \mbox{$\Delta R < 0.4 $} to the closest of the selected jets is then discarded.
We further identify jets as originating from a $b$-quark if their angular distance to
a $B$-hadron satisfies $\Delta R < 0.3$ and impose a $p_T$-dependent $b$-tagging
probability as described in Ref.~\cite{ATLAS:2011hfa}. This corresponds to an average
efficiency of 70\% in the case of $t\bar{t}$ events.

\subsection{Leptonic monotops}\label{sec:selection_lepton}
The preselection of events possibly containing a leptonically decaying
monotop signal has been directly designed from the expected final-state particle content.
As such, we demand the presence of exactly one lepton
candidate with a transverse momentum $p_T^\ell > 30$~GeV and one $b$-jet with a
transverse momentum $p_T^b > 30$~GeV. To reflect the expectation that the
produced supersymmetric particles (and their decay products)
are largely invisible, any
event containing an extra jet with a transverse momentum $p_T^j$ such that
$p_T^j > \mathrm{min}(p_T^b, 40 \GeV)$ is discarded.
After these basic requirements, a number of additional selection steps have been
implemented to increase the sensitivity $s$ of the analysis to the signal, where
$s=S/\sqrt{S+B}$ in which $S$ and $B$ are the number of signal and
background events passing all selection criteria, respectively. We define in
this way two signal regions, SRL1 and SRL2, dedicated to
the high and low mass regions of the superparticle parameter space, respectively.

Starting with the signal region SRL1 more sensitive to high mass setups, we impose that the
missing transverse momentum $\mathbf{p}_T^{\mathrm{miss}}$ in the event,
determined from the vector sum of the transverse momenta of all visible
final-state particles, has a magnitude $E_T^{\mathrm{miss}} > 150 \GeV$.
The orientation of the missing transverse momentum with respect to
the identified lepton is constrained by imposing a
minimum value to the $W$-boson transverse mass,
\be\label{eq:mtw}
  m_T^W = \sqrt{2 p_T^\ell E_T^{\rm{miss}} \Big[1 -
    \cos\big(\Delta \phi(\ell,\mathbf{p}_T^{\mathrm{miss}})\big)\Big]} \ ,
\ee
where $\Delta \phi(\ell,\mathbf{p}_T^{\mathrm{miss}})$ is the difference in azimuthal angle between the lepton and
the missing transverse momentum. We require selected events to satisfy
$m_T^W > 120 \GeV$, since when the missing transverse momentum in the event
originates solely from the leptonic decay of a $W$-boson, the $m_T^W$
distribution peaks at a lower value than when it finds its source both in a
$W$-boson decay and in a pair of invisible particles (like in the signal case).
This last selection ensures that the non-simulated QCD multijet background is
negligible~\cite{Aad:2012twa,Chatrchyan:2012bd}.

The second signal region SRL2 has been optimized for lower mass scenarios where
$E_T^{\mathrm{miss}}$ is typically very small due to a low neutralino mass.
Instead of constraining the individual quantities $E_T^{\mathrm{miss}}$ and
$m_T^W$, we select events satisfying
\be
  E_T^{\mathrm{miss}} + m_T^W > 220 \GeV.
\ee
In doing so, signal events with low values of $E_T^{\mathrm{miss}}$ are retained
by the selection process provided they have a suitably large value of $m_T^W$.
This still ensures that the QCD multijet background contributions are
small~\cite{Radics:2010iqa}.

In both search strategies, the following selection criteria are imposed.
Firstly, in order to reduce the number of background events in which the identified lepton and $b$-jet do not originate from a single top quark, a restriction on the invariant mass of the lepton plus $b$-jet system
is imposed,
 \be
  m_{b\ell} = \sqrt{(p^b + p^\ell)^2}<150 \GeV.
\ee
This leads to a reduction in background contributions from semi-leptonically decaying $t\bar{t}$ and $s$-channel single-top events in which one of the $b$-jets has not been identified.
Secondly, we enforce a minimum value to the invariant mass of the monotop system,
or equivalently to the invariant mass of the system comprised of the
missing transverse momentum, the identified lepton and $b$-jet,
\be
   m \left(\mathbf{p}_T^{\mathrm{miss}},\ell,b \right)
    = \sqrt{(\mathbf{p}_T^{\mathrm{miss}}+p^\ell+p^b)^2} > 700 \GeV.
\ee

\subsection{Hadronic monotops}\label{sec:selection_hadron}
In the hadronic case, final states related to the production
of a pair of strong superpartners together with a top quark are comprised of one
heavy-flavour and two lighter jets associated with the top decay,
as well as missing energy and extra soft objects arising from the decays of
the produced superparticles. We therefore preselect events that contain no
candidate leptons and exactly one $b$-jet with $p_T^b>30$~GeV. We however
demand exactly three light jets and not two, this requirement being found
to slightly increase the analysis sensitivity.

We design two search
strategies that we denote by SRH1 and SRH2. The former aims to be sensitive to
scenarios with higher superparticle masses and the latter to lower mass cases,
and we require the event missing energy to satisfy
$E_T^{\mathrm{miss}}> 200 \GeV$ and
$E_T^{\mathrm{miss}}> 150 \GeV$
for the SRH1 and SRH2 regions, respectively. While an even looser missing
energy selection might
increase the sensitivity in the SRH2 case, this would no longer ensure
a sufficient control of the non-simulated QCD multijet background and
furthermore not be sensible in the context of event triggers. To improve the
trigger efficiency associated with the SRH2 region,
we further require the hardest non $b$-tagged jet in each event
to fulfill $p_T^{j1} > 80 \GeV$, such that a trigger based on the selection of a
hard jet in association with missing transverse energy may be used. In contrast,
no selection on the hardest jet is imposed for the SRH1 region since
triggers based on the amount of missing energy only can be used.

A number of selection criterion are imposed for both strategies to improve the sensitivity of the analysis.  Firstly, the invariant mass of a light dijet system, $m_{jj}$, must be consistent with the mass of the $W$-boson,
\be
50 \GeV < m_{jj} < 100 \GeV,
\ee
where the pair of light jets is chosen such that the quantity $|m_W-m_{jj}|$ is minimized.  This pair of light-jets is then combined with the $b$-tagged jet to fully reconstruct the hadronically decaying top quark,
the resulting system being constrained to have an invariant mass in the range
\be
100 \GeV < m_{bjj} < 200 \GeV.
\ee
This eliminates a large number of background events which do not contain a hadronically decaying top quark.  In particular, it leads to a significant reduction of the $W$- and $\gamma^*/Z$-boson plus jet contributions.

Next, several restrictions are applied based on the kinematic configuration of the events.  The azimuthal angle between the missing transverse momentum and both the $b$-tagged and hardest non $b$-tagged jet in the event are required to be suitably large,
\be
  \Delta \phi \left(\mathbf{p}_T^{\mathrm{miss}}, \mathbf{p}^{j1}\right) > 0.6
  \quad\mathrm{and}\quad
  \Delta \phi \left(\mathbf{p}_T^{\mathrm{miss}}, \mathbf{p}^{b}\right) > 0.6\ .
\ee
These selection criteria are designed to rejected events in which the missing transverse energy originates from the mismeasurement of jets or semi-leptonic decays of heavy-fla\-vour hadrons.  Including these requirements is also expected to reduce background contributions originating from QCD multijet events with large instrumental missing transverse energy.  Finally to reflect the topology of signal events, the reconstructed top quark must be well separated from the missing transverse momentum with the difference in a\-zi\-mu\-thal angle exceeding
\be
\Delta \phi \left(\mathbf{p}_T^{\mathrm{miss}}, \mathbf{p}^t\right) > 1.8\ .
\ee

\section{Investigating compressed supersymmetric spectra with monotops} \label{sec:results}
\subsection{Leptonic monotops}\label{sec:results_lepton}
\begin{figure*}
  \centering
  \includegraphics[width=.35\textwidth]{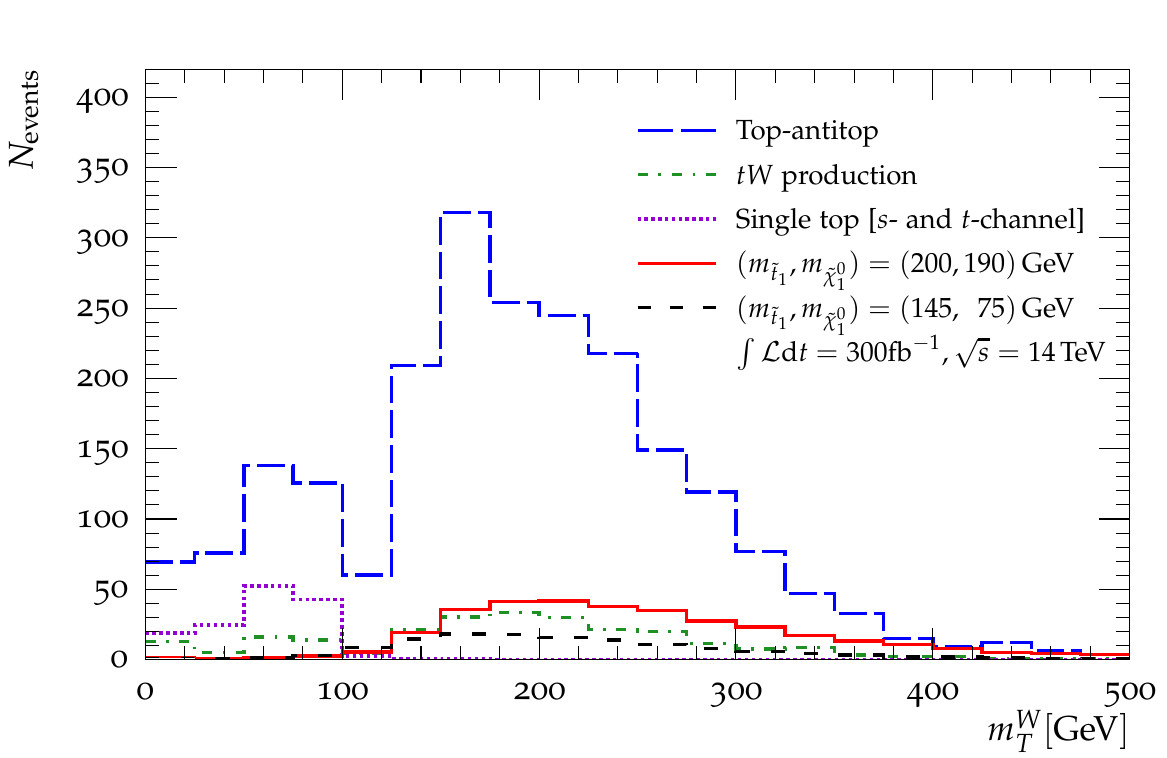}
  \includegraphics[width=.35\textwidth]{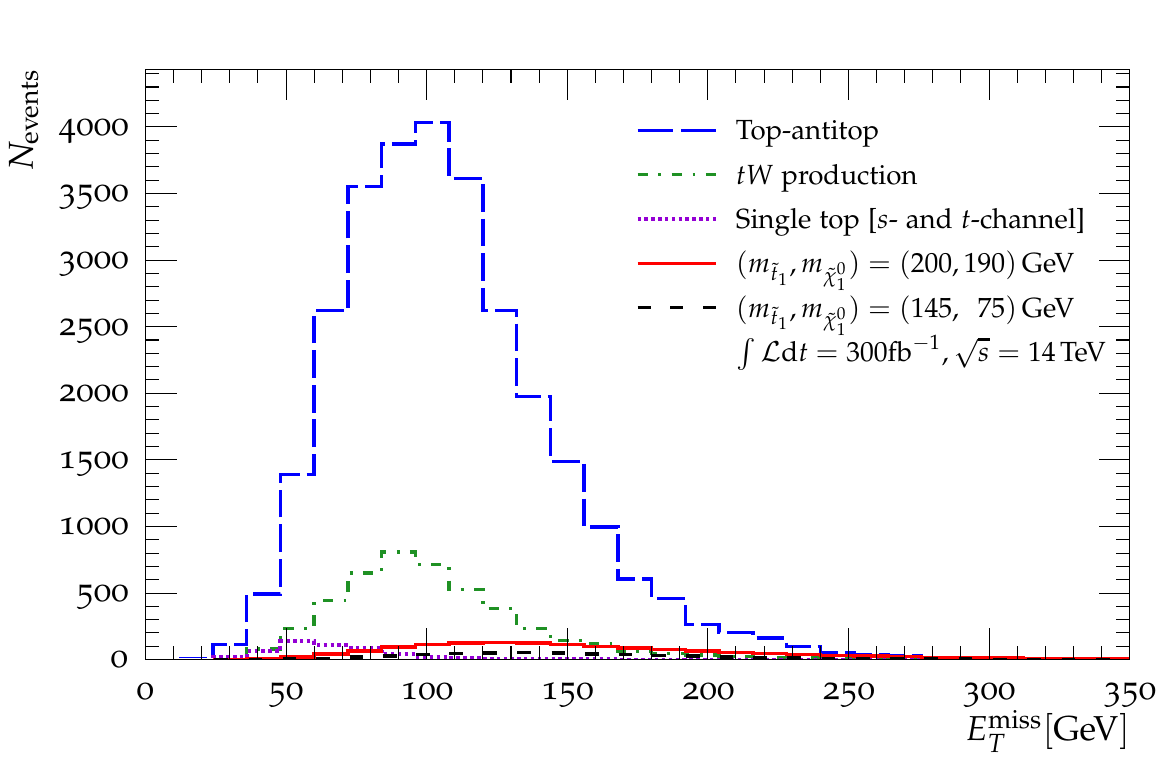}
  \caption{Distribution of the $W$-boson transverse mass $m_T^W$ (left) and
    of the missing transverse momentum $E_T^{\mathrm{miss}}$ (right),
    normalized to 300~fb$^{-1}$ of LHC collisions at a centre-of-mass
    energy of \mbox{$\sqrt{s} = 14$~TeV}. We present results for the dominant
    background contributions after all selection criteria defining the SRL1
    region have been applied, except \mbox{$m_T^W > 120 \GeV$} (left) and
    \mbox{$E_T^{\mathrm{miss}} > 150 \GeV$} (right). Also shown are the spectra
    for the example signal scenarios of Section~\ref{sec:simusignal}.}
  \label{fig:leptonic_exem}
\end{figure*}

\begin{figure*}
    \centering
    \includegraphics[width=.35\textwidth]{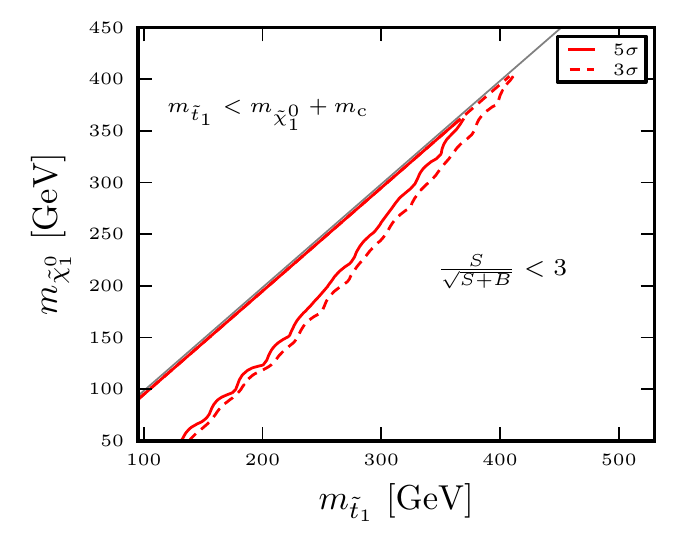}
    \includegraphics[width=.35\textwidth]{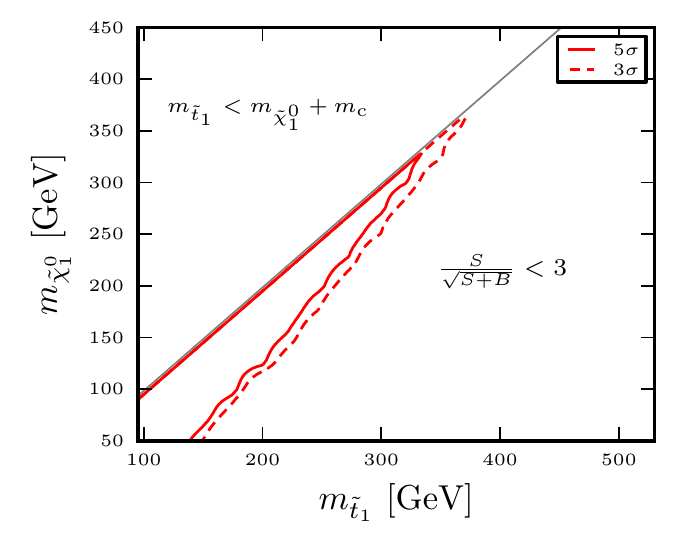}
  \caption{LHC sensitivity to a leptonically decaying monotop signal induced
    by a compressed supersymmetric scenario, adopting either the SRL1 (left) or
    SRL2 (right) search strategy. Results are shown in the
    $(m_{\tilde{t}_1},m_{\tilde{\chi}^0_1})$ plane for  
    $(\tilde{t}_1, t, \tilde{g})$ production in scenarios featuring 
    $m_{\tilde{t}_1}=m_{\tilde{g}}$.  Results are based on the simulation of 300~fb$^{-1}$ of LHC collisions at a
    centre-of-mass energy of $\sqrt{s} = 14$~TeV.}
  \label{fig:leptonic_scan_comp}
\end{figure*}

The numbers of events populating both leptonic monotop signal regions defined in Section~\ref{sec:selection_lepton} are listed in the third and fourth columns of Table~\ref{tab:sigma}, separately for the different background contributions and for the two compressed spectra scenarios mentioned in Section~\ref{sec:simusignal}. We recall that for the first of these, we have adopted a scenario in which $(m_{\tilde{t}_1} \!=\! m_{\tilde{g}},m_{\tilde{\chi}_1^0})= (200,190)$~GeV as a representative high mass setup with top squark, top quark and gluino production while we have chosen $(m_{\tilde{t}_1},m_{\tilde{\chi}_1^0})= (145,75)$~GeV as an example low mass scenario where the gaugino produced is instead the lightest neutralino.  The SRL1 analysis strategy is illustrated in Figure~\ref{fig:leptonic_exem} where we present the $m_T^W$ (left panel) and $E_T^{\rm{miss}}$ (right panel) distributions after applying all SRL1 selection requirements, except $m_T^W> 120 \GeV$ and $E_T^{\rm{miss}}>150$~GeV respectively. We show results for the two considered signal scenarios and the prevailing components of the Standard Model background, \ie~for $t\bar{t}$, $tW$ and single-top (in the $s$- and $t$-channel) production.

\begin{figure*}
    \centering
    \includegraphics[width=.35\textwidth]{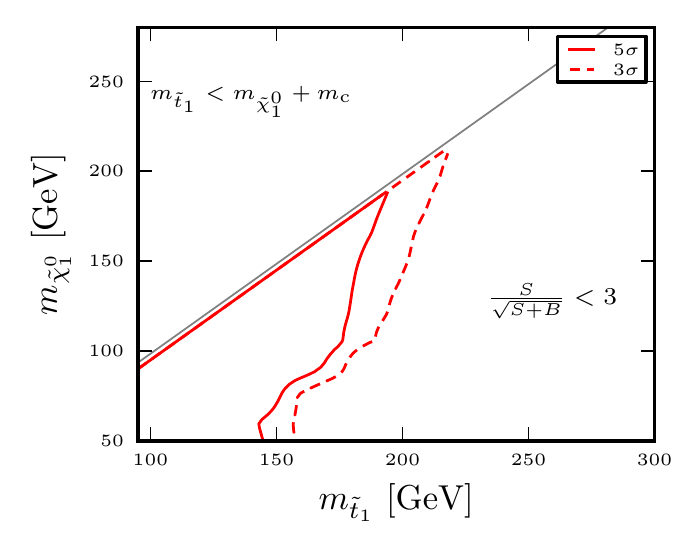}
    \includegraphics[width=.35\textwidth]{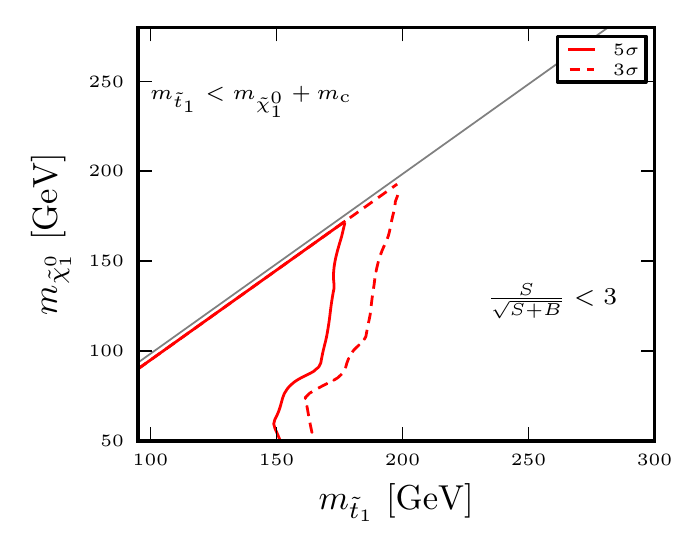}
  \caption{LHC sensitivity to a leptonically decaying monotop signal induced
    by a compressed supersymmetric scenario, adopting either the SRL1 (left) or
    SRL2 (right) search strategy. Results are shown in the
    $(m_{\tilde{t}_1},m_{\tilde{\chi}^0_1})$ plane for $(\tilde{t}_1, t, \tilde{\chi}_1^0)$ production with 300~fb$^{-1}$ of LHC collisions at a
    centre-of-mass energy of $\sqrt{s} = 14$~TeV.}
  \label{fig:leptonic_scan_higgsino}
\end{figure*}

The $m_T^W$ distribution for the background exhibits a peak in the region
$m_T^W \simeq 80 \GeV$, which corresponds to events
in which both the lepton and all the missing transverse
momentum originate from a $W$-boson decay. In contrast, both signal distributions
feature a suppression for $m_T^W<120$~GeV, which motivates
the $m_T^W$ selection criterion of the SRL1 strategy. In spite of the large number
of remaining background events, the sensitivity of the SRL1 analysis
to the high mass $(\tilde{t}_1, t, \tilde{g})$ and low mass  
$(\tilde{t}_1, t, \tilde{\chi}_1^0)$ signal scenarios
reaches $36\sigma$ and $14\sigma$ respectively. The SRL1 search strategy having been
designed to probe higher mass spectra, is by construction
less sensitive to lower neutralino masses where the smaller
neutralino mass yields comparatively less $E_T^{\rm{miss}}$, as depicted
on the right panel of Figure~\ref{fig:leptonic_exem}.
This drop in the sensitivity is 
alleviated through the inclusion of the SRL2 analysis strategy.  Accordingly, SRL2 exhibits a reduced sensitivity to the high mass benchmark point of $29\sigma$.  The sensitivity to the lower mass $(\tilde{t}_1, t, \tilde{\chi}_1^0)$ example scenario is also slightly reduced to $13 \sigma$.  However, it has been confirmed that in even less compressed scenarios the sensitivity of the SRL2 analysis exceeds that of the SRL1 analysis, a feature that is evident in the discovery bounds presented in Figure~\ref{fig:leptonic_scan_higgsino}.

\begin{figure*}
  \centering
  \includegraphics[width=.35\textwidth]{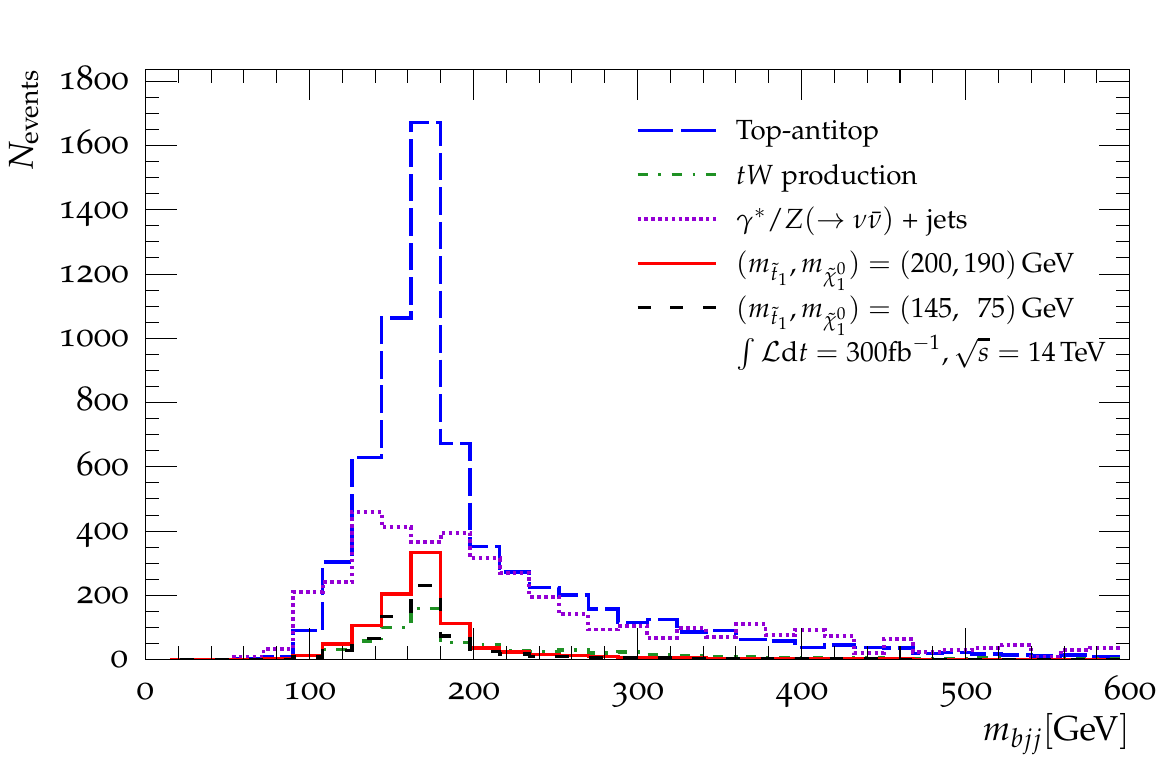}
  \caption{Distribution of the invariant mass of the reconstructed top quark
    $m_{bjj}$, normalized to 300~fb$^{-1}$ of LHC collisions at a centre-of-mass
    energy of $\sqrt{s} = 14$~TeV. We present
    results for the dominant background contributions after all selection
    criteria defining the SRH2 region have been applied, except
    \mbox{$m_{bjj}\in[100,200] \GeV$}. Also shown are the spectra
    for the example signal scenarios of Section~\ref{sec:simusignal}.}
  \label{fig:hadronic}
\end{figure*}

To study more extensively the LHC sensitivity to different supersymmetric
scenarios featuring small mass gaps among the lightest superpartners,
we scan in the $(m_{\tilde{t}_1},m_{\tilde{\chi}_1^0})$ plane, enforcing
$m_{\tilde{t}_1} = m_{\tilde{g}}$ for the $(\tilde{t}_1,t, \tilde{g})$ production signal scenario, and derive contours
corresponding to different observation boundaries. The $5\sigma$ and
$3\sigma$ regions for top quark, top squark and gluino production are respectively shown by solid and dashed red lines in
Figure~\ref{fig:leptonic_scan_comp} where we present the independent
contributions of the SRL1 (left panel) and SRL2 (right panel) search strategies.  Equivalent boundaries are shown in Figure~\ref{fig:leptonic_scan_higgsino} for the case of direct neutralino production in association with a top and stop.
As a result of the design, the SRL1 analysis is found to be more sensitive to higher mass setups for both signal scenarios.  The SRL2 search strategy is more sensitive to scenarios featuring smaller
superpartner masses and possibly less compressed spectra, as can be seen for both $(\tilde{t}_1,t, \tilde{g})$ and $(\tilde{t}_1,t, \tilde{\chi}_1^0)$ production with the latter exhibiting a more significant improvement in the less compressed regions of parameter space.

\subsection{Hadronic monotops}\label{sec:results_hadron}

We again focus on the high mass and low mass example scenarios of
Section~\ref{sec:simusignal} and apply the hadronic monotop selection
requirements outlined in
Section~\ref{sec:selection_hadron}. The number of signal events populating the
SRH1 and SRH2 regions are presented in the fifth and sixth columns of
Table~\ref{tab:sigma}, together with the
different background contributions. The results indicate that the Standard Model
background is largely comprised of events originating from $t \bar{t}$,
$\gamma^*/Z$-boson plus jets, $tW$ and $W$-boson plus light-jet production for
both search regions.

Our hadronic monotop selection strategy is illustrated in
Figure~\ref{fig:hadronic} where we show the distribution in the invariant-mass
of the reconstructed top quark after applying all SRH2 requirements, except the
one on $m_{bjj}$, for the two considered
signal scenarios and the dominant background sources\footnote{In principle, the
subdominant $W$-boson plus light-jet results should also be represented.
However, only a very small fraction of the $\sim10^8$ simulated events have
passed all selection
criteria, so that after normalizing to the large associated total
cross section and an integrated luminosity of 300~fb$^{-1}$
the resulting statistical uncertainty is
important. The $W$-boson plus light-jets curve has therefore been omitted.}.

Imposing the constraint
\mbox{$m_{bjj}\in[100,200]$~GeV} will retain the majority of the signal
events while reducing the number of background events, particularly in the case
of $\gamma^*/Z$ plus jets production for which the distribution does not peak
significantly at the top mass. As such, after applying this final selection
criteria the sensitivity of the SRH2 strategy to the high mass $(\tilde{t}_1,t, \tilde{g})$ and low mass $(\tilde{t}_1,t, \tilde{\chi}_1^0)$ signal
benchmark points is found to be
$38\sigma$ and $26\sigma$ respectively. The SRH1 strategy being in contrast
dedicated to higher mass setups, its sensitivity to the high mass scenario
is found to be significantly improved and reaches $45\sigma$, whilst it drops slightly
to $25\sigma$ for the low mass example.

\begin{figure*}
  \centering
  \includegraphics[width=.35\textwidth]{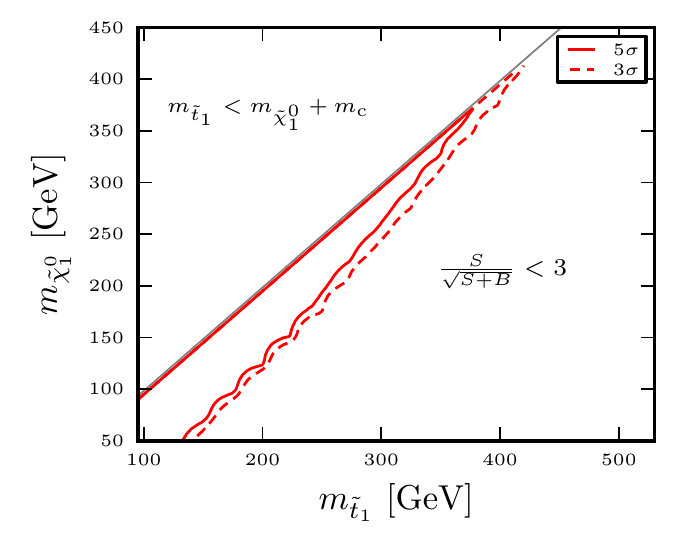}
  \includegraphics[width=.35\textwidth]{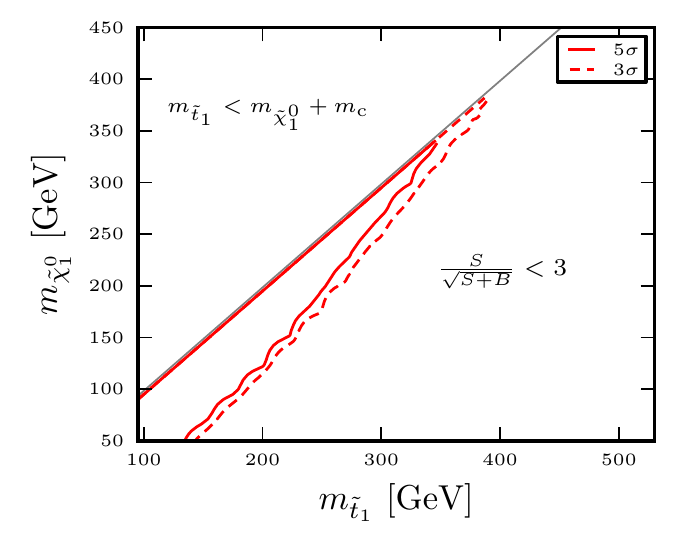}
  \caption{Same as Figure~\ref{fig:leptonic_scan_comp} but for the SRH1 (left)
    and SRH2 (right) search strategies.}
  \label{fig:hadronic_scan_comp}
\end{figure*}

\begin{figure*}
  \centering
  \includegraphics[width=.35\textwidth]{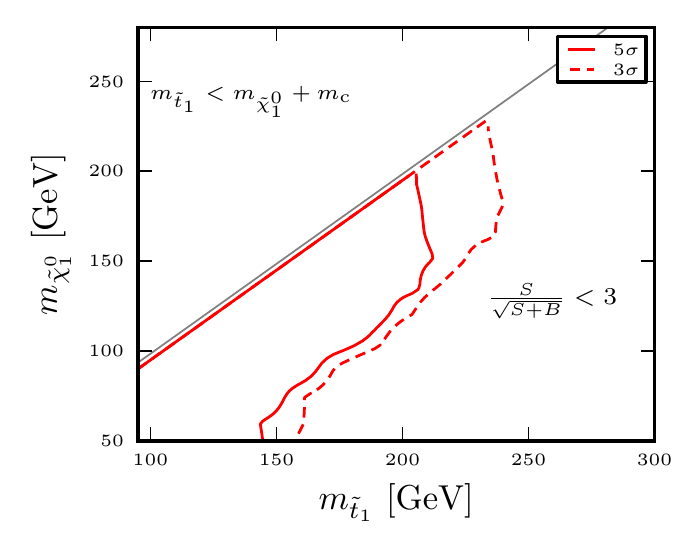}
  \includegraphics[width=.35\textwidth]{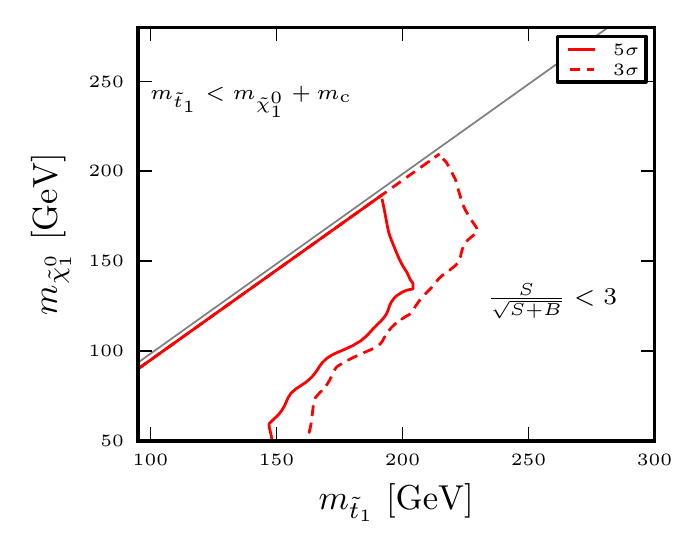}
  \caption{Same as Figure~\ref{fig:leptonic_scan_higgsino} but for the SRH1 (left)
    and SRH2 (right) search strategies.}
  \label{fig:hadronic_scan_higgsino}
\end{figure*}

As in Section~\ref{sec:results_lepton}, we perform a scan in the $(m_{\tilde{t}_1},
m_{\tilde{\chi}_1^0})$ plane, with the equality $m_{\tilde{t}_1} = m_{\tilde{g}}$
enforced for the case of $(\tilde{t}_1,t, \tilde{g})$ production. The results are given in Figures~\ref{fig:hadronic_scan_comp} and~\ref{fig:hadronic_scan_higgsino} where
we show the $5\sigma$ and $3\sigma$ contours found after applying the SRH1 (left
panel) and SRH2 (right panel) search strategies for $(\tilde{t}_1,t, \tilde{g})$ and $(\tilde{t}_1,t, \tilde{\chi}_1^0)$ production respectively. By design, the
SRH1 analysis presents an enhanced sensitivity to compressed scenarios featuring
large superparticle masses while the SRH2 strategy is instead more tuned to situations exhibiting
smaller superparticle masses, with a possibly less compressed spectrum. The reach of SRH2 improves over the that of SRH1 in the low mass region of the parameter space for the case of $(\tilde{t}_1,t, \tilde{\chi}_1^0)$ production.   However, no significant extension of the observation boundaries is seen in the case of $(\tilde{t}_1,t, \tilde{g})$ production in Figure~\ref{fig:hadronic_scan_comp}.  

Finally, we note that the contours derived by considering hadronic monotop decays in $(\tilde{t}_1,t, \tilde{g})$ production exceed the limits set by the leptonic monotop search strategies in all regions of the $(m_{\tilde{t}_1}, m_{\tilde{\chi}_1^0})$ plane.  For the case of $(\tilde{t}_1,t, \tilde{\chi}_1^0)$ production, hadronic monotop decays provide the most extensive reach in high mass scenarios, while considering leptonic decays leads to more stringent limits in low mass and less compressed regions of the $(m_{\tilde{t}_1}, m_{\tilde{\chi}_1^0})$ plane.

\subsection{Comparison to existing bounds}

For the signal scenario in which a gluino is produced in association with the top squark and top quark, the search strategies presented here have the capability of discovering a significant region of the $(m_{\tilde{t}_1}, m_{\chi_1^0})$ plane.  However, these scenarios are in fact already excluded.  By requiring $m_{\tilde{g}}=m_{\tilde{t}_1}$ the signal is subject to constraints derived from direct LHC searches for light gluinos~\cite{Aad:2014wea,Chatrchyan:2014lfa}.  These rule out at the 95\% confidence level the existence of gluinos with mass less than $\mathcal{O}(600)$~GeV in highly compressed scenarios. Even with 300~fb$^{-1}$ of LHC collisions at a centre-of-mass energy of $\sqrt{s}=$14~TeV, our monotop based search strategy does not have comparable sensitivity to these higher mass gluino scenarios.

\begin{figure*}
  \centering
  \includegraphics[width=0.35\textwidth]{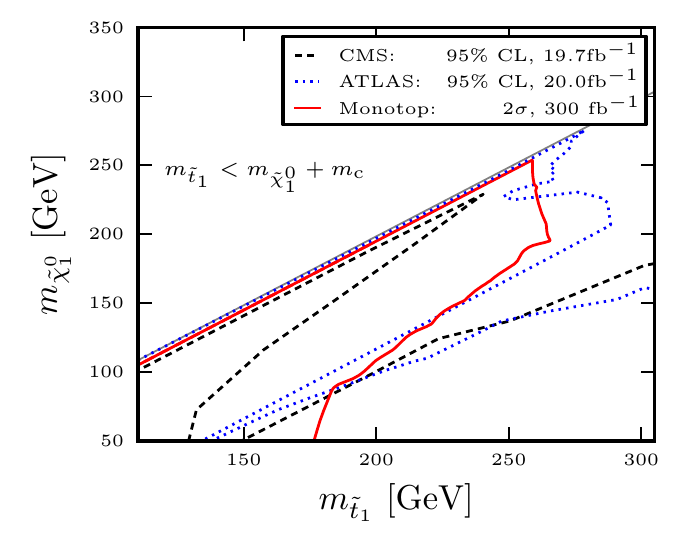}
  \includegraphics[width=0.35\textwidth]{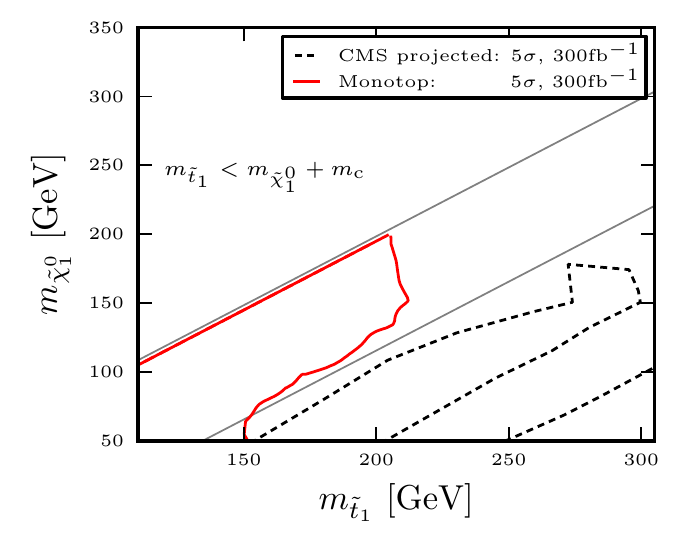}
  \caption{Left: Comparison of the current 95\% CL exclusion boundaries set by the ATLAS and CMS collaborations with $\mathcal{O}(20)~\mathrm{fb}^{-1}$ of data recorded at $\sqrt{s}=8$~TeV and the $2\sigma$ sensitivity of the LHC to a monotop signal arising from $(\tilde{t}_1, t, \tilde{\chi}_1^0)$ production for 300~fb$^{-1}$ of LHC collisions at a
    centre-of-mass energy of $\sqrt{s} = 14$~TeV.  Right: Comparison of the $5 \sigma$ sensitivity of the LHC to hadonically and leptonically decaying monotop signals arising from $(\tilde{t}_1, t, \tilde{\chi}_1^0)$ production and the extrapolated $5 \sigma$ discovery bound derived from a CMS search for stops in events with electrons and muons~\cite{CMS:2013hda}. Both results correspond to 300~fb$^{-1}$ of LHC data at $\sqrt{s} = 14$~TeV. }
  \label{fig:combined_scan_higgsino}
\end{figure*}

For our second signal scenario in which a top squark and top quark are produced in association with the lightest neutralino, the gluino mass bounds are no longer applicable.  As such, we investigate whether the monotop based search strategy presented here can place competitive exclusion limits when compared with more traditional approaches that search for monojet events or make use of charm-flavour identification techniques.  To do so, we approximate the 95\% CL exclusion limit of our search strategy with the 2$\sigma$ discovery bound and plot this contour in the left-hand panel of Figure~\ref{fig:combined_scan_higgsino} for the combined hadronic monotop search strategies\footnote{We find in this case that the discovery bounds derived from studying hadronic monotop decays exceed those arising from leptonically decaying monotops searches in all regions of the parameter space.}.  Superimposed on Figure~\ref{fig:combined_scan_higgsino} are the the current 95\% CL exclusion limits set by the ATLAS~\cite{atlassusy} and CMS~\cite{CMS:2014yma} collaborations.
We observe that our monotop based search strategy can provide comparable exclusion bounds in the region with $m_{\tilde{t}_1}< m_{\tilde{\chi}_1^0}+m_b+m_W$ and exceed the existing limits in the region with $m_{\tilde{t}_1} \approx m_{\tilde{\chi}_1^0}+m_b+m_W$ and $m_{\tilde{t}_1} \lesssim 210$~GeV.  However, we note that our bounds make use of $300~\mathrm{fb}^{-1}$ of LHC data at $\sqrt{s}=14$~TeV while the existing limits are based on $\mathcal{O}(20)~\mathrm{fb}^{-1}$ of data recorded at $\sqrt{s}=8$~TeV.

The right-hand panel of Figure~\ref{fig:combined_scan_higgsino} instead shows a comparison between the $5\sigma$ discovery reach of our monotop analysis\footnote{We shown the result obtained by combining the hadronic and leptonic monotop search strategies in the most naive way, making use of the most sensitive search strategy at each parameter space point.  While a more sophisticated combination might further expand the observation boundaries, our approach provides a conservative estimate.} and the $5\sigma$ discovery boundary arising from a CMS search for top squarks in events with final-state electrons or muons~\cite{CMS:2013hda}, extrapolated to $300~\mathrm{fb}^{-1}$ of LHC data taken at $\sqrt{s}=14$~TeV~\cite{CMS:2013xfa}.  For the latter, we show the boundary corresponding to the less conservative scenario in which the uncertainty on the background is assumed to be dominated by the statistical precision.  Here we observe that the monotop analysis sets stronger limits in the region $m_{\tilde{t}_1} \approx m_{\tilde{\chi}_1^0}+m_b+m_W$ for $m_{\tilde{t}_1} \lesssim 150$~GeV.  However, the comparison is again not ideal given that extrapolated boundaries are not available for the search strategies which set the most stringent limits in the compressed regions of phase space with $\sqrt{s}=8$~TeV data.

\section{Conclusions}\label{sec:conclusions}
We have investigated the feasibility of using monotop probes to get a handle on
supersymmetric scenarios featuring a compressed spectrum. We have considered the
production of a pair of superparticles in association with a
top quark from proton-proton collisions at a centre-of-mass energy of 14~TeV.
The supersymmetric spectrum being compressed, both superpartners decay into
missing energy carried by the lightest supersymmetric particle and a collection
of objects too soft to be reconstructed.  The resulting new
physics signal consequently consists of a monotop signature.

Both the leptonic and hadronic decays of the top quark have been
investigated and two pairs of analysis strategies, respectively dedicated to
the low and high mass regions of the parameter space, have been designed.

We have shown that monotop signals arising from the production of a top squark, a top quark and a gluino in a compressed supersymmetric setup are in principle reachable with $5\sigma$ sensitivity at the future run II of the LHC with a luminosity of 300~fb$^{-1}$ in the case where the top squark and gluino masses are below 380~GeV.  However, we find the monotop based search strategy is not competitive with current bounds set by direct searches for light gluinos.

Additionally, we have studied the production of a top squark and top quark in association with the lightest neutralino.  In this case, $5 \sigma$ sensitivity is obtained for compressed scenarios with $m_{\tilde{t}_1} \lesssim 200$~GeV and also in the region $m_{\tilde{t}_1} \approx m_{\tilde{\chi}_1^0}+m_b+m_W$ for $m_{\tilde{t}_1} \lesssim 150$~GeV.  The latter region is not excluded by any existing extrapolations of current searches to 300~fb$^{-1}$ of $\sqrt{s}=14$~TeV data.

\begin{acknowledgements}
We are grateful to the organizers of the 2013 Les Houches
\textit{Physics at TeV Colliders} workshop where this work has been initiated,
as well as to Filip Moortgat for enlightening discussions.
he work of BF has been partially supported by the Theory-LHC France initiative
f the CNRS/IN2P3. AW and PR acknowledge the support of the European Union via 
MCNet, PITN-GA-2012-315877 and the Science and Technology Facilities Council.
\end{acknowledgements}

\bibliography{bibliography}

\begin{thebibliography}{10}
\providecommand{\url}[1]{\texttt{#1}}
\providecommand{\urlprefix}{URL }
\providecommand{\eprint}[2][]{\url{#2}}

\bibitem{Nilles:1983ge}
H.~P. Nilles, Phys.Rept. \textbf{110}, 1 (1984)

\bibitem{Haber:1984rc}
H.~E. Haber, G.~L. Kane, Phys.Rept. \textbf{117}, 75 (1985),
  \eprint{UM-HE-TH-83-17, SCIPP-85-47}

\bibitem{atlassusy}
{https://twiki.cern.ch/twiki/bin/view/AtlasPublic/ \\
  SupersymmetryPublicResults}

\bibitem{cmssusy}
{https://twiki.cern.ch/twiki/bin/view/CMSPublic/ \\ PhysicsResultsSUS}

\bibitem{atlasmet}
{https://twiki.cern.ch/twiki/bin/view/AtlasPublic/ \\
  MissingEtTriggerPublicResults}

\bibitem{cmsmet}
{Private Communication}

\bibitem{Alves:2010za}
D.~S. Alves, E.~Izaguirre, J.~G. Wacker, Phys.Lett. \textbf{B702}, 64 (2011),
  \eprint{1008.0407}

\bibitem{LeCompte:2011cn}
T.~J. LeCompte, S.~P. Martin, Phys.Rev. \textbf{D84}, 015004 (2011),
  \eprint{1105.4304}

\bibitem{LeCompte:2011fh}
T.~J. LeCompte, S.~P. Martin, Phys.Rev. \textbf{D85}, 035023 (2012),
  \eprint{1111.6897}

\bibitem{Alvarez:2012wf}
E.~Alvarez, Y.~Bai, JHEP \textbf{1208}, 003 (2012), \eprint{1204.5182}

\bibitem{Dreiner:2012gx}
H.~K. Dreiner, M.~Kramer, J.~Tattersall, Europhys.Lett. \textbf{99}, 61001
  (2012), \eprint{1207.1613}

\bibitem{Bhattacherjee:2012mz}
B.~Bhattacherjee, K.~Ghosh  (2012), \eprint{1207.6289}

\bibitem{Dreiner:2012sh}
H.~Dreiner, M.~Kramer, J.~Tattersall, Phys.Rev. \textbf{D87}, 3, 035006 (2013),
  \eprint{1211.4981}

\bibitem{Ghosh:2013qga}
D.~Ghosh, Phys.Rev. \textbf{D88}, 115013 (2013), \eprint{1308.0320}

\bibitem{Belanger:2013oka}
G.~Belanger, D.~Ghosh, R.~Godbole, et~al., Phys.Rev. \textbf{D89}, 015003
  (2014), \eprint{1308.6484}

\bibitem{Dutta:2013gga}
B.~Dutta, W.~Flanagan, A.~Gurrola, et~al.  (2013), \eprint{1312.1348}

\bibitem{Bhattacherjee:2013wna}
B.~Bhattacherjee, A.~Choudhury, K.~Ghosh, et~al., Phys.Rev. \textbf{D89},
  037702 (2014), \eprint{1308.1526}

\bibitem{Schwaller:2013baa}
P.~Schwaller, J.~Zurita, JHEP \textbf{1403}, 060 (2014), \eprint{1312.7350}

\bibitem{Deppisch:2014aga}
F.~F. Deppisch, N.~Desai, T.~E. Gonzalo, Front.Phys. \textbf{2}, 00027 (2014),
  \eprint{1403.2312}

\bibitem{Andrea:2011ws}
J.~Andrea, B.~Fuks, F.~Maltoni, Phys.Rev. \textbf{D84}, 074025 (2011),
  \eprint{1106.6199}

\bibitem{Kamenik:2011nb}
J.~F. Kamenik, J.~Zupan, Phys.Rev. \textbf{D84}, 111502 (2011),
  \eprint{1107.0623}

\bibitem{Wang:2011uxa}
J.~Wang, C.~S. Li, D.~Y. Shao, et~al., Phys.Rev. \textbf{D86}, 034008 (2012),
  \eprint{1109.5963}

\bibitem{Fuks:2012im}
B.~Fuks, Int.J.Mod.Phys. \textbf{A27}, 1230007 (2012), \eprint{1202.4769}

\bibitem{Alvarez:2013jqa}
E.~Alvarez, E.~C. Leskow, J.~Drobnak, et~al., Phys.Rev. \textbf{D89}, 014016
  (2014), \eprint{1310.7600}

\bibitem{Agram:2013wda}
J.-L. Agram, J.~Andrea, M.~Buttignol, et~al., Phys.Rev. \textbf{D89}, 014028
  (2014), \eprint{1311.6478}

\bibitem{Boucheneb:2014wza}
I.~Boucheneb, G.~Cacciapaglia, A.~Deandrea, et~al., JHEP \textbf{1501}, 017
  (2015), \eprint{1407.7529}

\bibitem{Aaltonen:2012ek}
T.~Aaltonen, et~al. (CDF), Phys.Rev.Lett. \textbf{108}, 201802 (2012),
  \eprint{1202.5653}

\bibitem{CMS:2014hba}
CMS  (2014), \eprint{CMS-PAS-B2G-12-022}

\bibitem{Khachatryan:2014uma}
V.~Khachatryan, et~al. (CMS)  (2014), \eprint{1410.1149}

\bibitem{Aad:2014wza}
G.~Aad, et~al. (ATLAS), Eur.Phys.J. \textbf{C75}, 2, 79 (2015),
  \eprint{1410.5404}

\bibitem{Bozzi:2007me}
G.~Bozzi, B.~Fuks, B.~Herrmann, et~al., Nucl.Phys. \textbf{B787}, 1 (2007),
  \eprint{0704.1826}

\bibitem{Fuks:2008ab}
B.~Fuks, B.~Herrmann, M.~Klasen, Nucl.Phys. \textbf{B810}, 266 (2009),
  \eprint{0808.1104}

\bibitem{Fuks:2011dg}
B.~Fuks, B.~Herrmann, M.~Klasen, Phys.Rev. \textbf{D86}, 015002 (2012),
  \eprint{1112.4838}

\bibitem{Christensen:2009jx}
N.~D. Christensen, P.~de~Aquino, C.~Degrande, et~al., Eur.Phys.J. \textbf{C71},
  1541 (2011), \eprint{0906.2474}

\bibitem{Duhr:2011se}
C.~Duhr, B.~Fuks, Comput.Phys.Commun. \textbf{182}, 2404 (2011),
  \eprint{1102.4191}

\bibitem{Degrande:2011ua}
C.~Degrande, C.~Duhr, B.~Fuks, et~al., Comput.Phys.Commun. \textbf{183}, 1201
  (2012), \eprint{1108.2040}

\bibitem{Alloul:2013bka}
A.~Alloul, N.~D. Christensen, C.~Degrande, et~al., Comput.Phys.Commun.
  \textbf{185}, 2250 (2014), \eprint{1310.1921}

\bibitem{Alwall:2014hca}
J.~Alwall, R.~Frederix, S.~Frixione, et~al., JHEP \textbf{1407}, 079 (2014)

\bibitem{Pumplin:2002vw}
J.~Pumplin, D.~Stump, J.~Huston, et~al., JHEP \textbf{0207}, 012 (2002),
  \eprint{hep-ph/0201195}

\bibitem{Bahr:2008pv}
M.~Bahr, S.~Gieseke, M.~Gigg, et~al., Eur.Phys.J. \textbf{C58}, 639 (2008),
  \eprint{0803.0883}

\bibitem{Bellm:2013lba}
J.~Bellm, S.~Gieseke, D.~Grellscheid, et~al.  (2013), \eprint{1310.6877}

\bibitem{Degrande:2014sta}
C.~Degrande, B.~Fuks, V.~Hirschi, et~al.  (2014), \eprint{1412.5589}

\bibitem{Nason:2004rx}
P.~Nason, JHEP \textbf{0411}, 040 (2004), \eprint{hep-ph/0409146}

\bibitem{Frixione:2007vw}
S.~Frixione, P.~Nason, C.~Oleari, JHEP \textbf{0711}, 070 (2007),
  \eprint{0709.2092}

\bibitem{Alioli:2010xd}
S.~Alioli, P.~Nason, C.~Oleari, et~al., JHEP \textbf{1006}, 043 (2010),
  \eprint{1002.2581}

\bibitem{Frixione:2007nw}
S.~Frixione, P.~Nason, G.~Ridolfi, JHEP \textbf{0709}, 126 (2007),
  \eprint{0707.3088}

\bibitem{Hamilton:2008pd}
K.~Hamilton, P.~Richardson, J.~Tully, JHEP \textbf{0810}, 015 (2008),
  \eprint{0806.0290}

\bibitem{Alioli:2009je}
S.~Alioli, P.~Nason, C.~Oleari, et~al., JHEP \textbf{0909}, 111 (2009),
  \eprint{0907.4076}

\bibitem{Re:2010bp}
E.~Re, Eur.Phys.J. \textbf{C71}, 1547 (2011), \eprint{1009.2450}

\bibitem{Frixione:2008yi}
S.~Frixione, E.~Laenen, P.~Motylinski, et~al., JHEP \textbf{0807}, 029 (2008),
  \eprint{0805.3067}

\bibitem{Gleisberg:2003xi}
T.~Gleisberg, S.~Hoeche, F.~Krauss, et~al., JHEP \textbf{0402}, 056 (2004),
  \eprint{hep-ph/0311263}

\bibitem{Gleisberg:2008ta}
T.~Gleisberg, S.~Hoeche, F.~Krauss, et~al., JHEP \textbf{0902}, 007 (2009),
  \eprint{0811.4622}

\bibitem{Hamilton:2010wh}
K.~Hamilton, P.~Nason, JHEP \textbf{1006}, 039 (2010), \eprint{1004.1764}

\bibitem{Hoche:2010kg}
S.~Hoche, F.~Krauss, M.~Schonherr, et~al., JHEP \textbf{1108}, 123 (2011),
  \eprint{1009.1127}

\bibitem{Hamilton:2010mb}
K.~Hamilton, JHEP \textbf{1101}, 009 (2011), \eprint{1009.5391}

\bibitem{Aad:2012ux}
G.~Aad, et~al. (ATLAS), Phys.Lett. \textbf{B717}, 330 (2012),
  \eprint{1205.3130}

\bibitem{Cacciari:2008gp}
M.~Cacciari, G.~P. Salam, G.~Soyez, JHEP \textbf{0804}, 063 (2008),
  \eprint{0802.1189}

\bibitem{Cacciari:2011ma}
M.~Cacciari, G.~P. Salam, G.~Soyez, Eur.Phys.J. \textbf{C72}, 1896 (2012),
  \eprint{1111.6097}

\bibitem{ATLAS:2011hfa}
ATLAS  (2011), \eprint{ATLAS-CONF-2011-089, ATLAS-COM-CONF-2011-082}

\bibitem{Aad:2012twa}
G.~Aad, et~al. (ATLAS), Eur.Phys.J. \textbf{C72}, 2173 (2012),
  \eprint{1208.1390}

\bibitem{Chatrchyan:2012bd}
S.~Chatrchyan, et~al. (CMS), Eur.Phys.J. \textbf{C73}, 2283 (2013),
  \eprint{1210.7544}

\bibitem{Radics:2010iqa}
B.~Radics \eprint{CERN-THESIS-2010-237, BONN-IR-2010-14}

\bibitem{Aad:2014wea}
G.~Aad, et~al. (ATLAS Collaboration)  (2014), \eprint{1405.7875}

\bibitem{Chatrchyan:2014lfa}
S.~Chatrchyan, et~al. (CMS Collaboration), JHEP \textbf{1406}, 055 (2014),
  \eprint{1402.4770}

\bibitem{CMS:2013hda}
CMS  (2013), \eprint{CMS-PAS-SUS-13-011}

\bibitem{CMS:2014yma}
C.~Collaboration (CMS)  (2014)

\bibitem{CMS:2013xfa}
CMS  (2013), \eprint{1307.7135}

\end{thebibliography}

\end{document}